\documentclass[12pt]{article}
\usepackage{graphicx}
\begin{document}
\begin{center}
{\large\bf Stochastic String Topography and Multivacua} \vskip 0.3
true in {\large J. W. Moffat} \vskip 0.3 true in {\it The
Perimeter Institute for Theoretical Physics, Waterloo, Ontario,
N2J 2W9, Canada} \vskip 0.3 true in and \vskip 0.3 true in {\it
Department of Physics, University of Waterloo, Waterloo, Ontario
N2L 3G1, Canada}
\end{center}
\begin{abstract}%
The suggestion that there exist causally disconnected universes or
sub-universes to explain the values of physical parameters such as
the cosmological constant is discussed. A statistical model of the
string landscape/topography is formulated using a stochastic
Langevin equation for string and supergravity potentials. A
Focker-Planck equation for the probability density of
superpotentials is derived and the possible non-supersymmetric
multivacua describing string/M-theory topography are investigated.
The stochastic fluctuations of the superpotentials and their associated
vacuum states can possibly lead to a small positive cosmological constant.
\end{abstract}
\vskip 0.2 true in

e-mail: jmoffat@perimeterinstitute.ca


\section{Introduction}

There has for many active physicists been a disturbing increase in
the attention paid to the notion that we need a large, and
possibly infinite, number of universes that make up a multiverse,
${\cal M}$. These universes are causally disconnected and
therefore no observations can ever be performed to detect them.
Much of the argument evolving from the multiverse point of view
stems from various variations of the anthropic
principle~\cite{Carter,Rees,Barrow}.

There have been two developments in physics that have been
triggers for well-respected physicists to take such a serious
step. The first is the egregious disagreement between na\"ive
quantum field theory calculations of the cosmological constant
$\Lambda$ and the observations of the CMB and supernovae SNIa
~\cite{Perlmutter,Spergel}. The cosmological constant was
introduced by Einstein in his theory of gravity to stabilize the
universe through the repulsive force it causes in his field
equations. This constant came and went in the development of
cosmology, but has now made a triumphant return due to the
remarkable discovery that the expansion of the universe appears to
be accelerating.

The cause for this acceleration could be the existence of dark
energy, transparent to electromagnetic radiation, which can be
identified as a smooth background of vacuum energy density. This
identification is supported by WMAP and type SNIa supernova
data~\cite{Perlmutter,Spergel}, for in the equation of state for a
perfect fluid $p=w\rho$ the parameter $w$ is measured to be
$w=-1.02\pm 0.5$ and for the cosmological constant $w=-1$.
However, the consensus opinion in the physics community is that
all attempts have failed to explain why the na\"ive quantum field
theory calculation of the cosmological constant $\Lambda$ is $122$
orders of magnitude larger than the ``observed'' value, $\rho_{\rm
vac}\sim (10^{-3}\,{\rm eV})^4$, where $\Lambda=\rho_{\rm
vac}/M_P^{*2}$, $\rho_{\rm vac}$ is the vacuum energy density and
$M_P^*$ is the reduced Planck mass~\cite{Weinberg}. To make
matters worse, there is the preposterous outcome of the observed
acceleration of the universe that the dominance of the
cosmological constant in the evolution of the universe should have
happened within our present epoch. This ``coincidence'' has also
defied any simple explanation.

The second motivation is the observation that string theory must
have a very large number (``googleplex'') of vacuum
states~\cite{Susskind}. This has been suspected (known) for a long
time but certain string aficionados have emphasized this
circumstance in the recent
literature~\cite{Susskind,Douglas,Banks,Donoghue}, and there is less
attempt to ignore or deny that this phenomenon will play a
fundamental role in the development of string theory in the
future. Of course, string theory has always suffered from the lack
of a unique way to compactify the 10/11-dimensional theory to the
4 known spacetime dimensions. There is an infinite
number of ways to perform such a compactification. The problem
lies in the moduli fields that are ubiquitous in string theory and
that reliable calculations in string theory can only be performed
when the theory is supersymmetric. Recently, it has been proposed
that string theories can exist in dimensions higher than the
critical 10/11 dimensions~\cite{Hellerman}, in which case the lack
of uniqueness of string theory clearly increases enormously.

The original motivations for string theory were twofold: 1) It
would provide a consistent way to combine quantum theory with
gravity, 2) It could lead to a unification of the known four
forces of Nature. It is often stated by string theorists that
string theory is ``the only way to unify Einstein's GR with
quantum theory.'' This supposition has never been proved and
alternative ways of unifying the two pillars of modern physics
have been pursued over several years, and others will no doubt
make their appearance in future~\cite{Smolin}.

The second motivation is the desire of particle physicists to seek
a unification of the known forces of nature. This reductionist
approach could hopefully lead to an understanding of the many free
parameters in the Standard Model of particle physics and our
present (incomplete) model of cosmology \footnote{The number of
parameters in the minimum supersymmetric standard model (MSSM) is
of order 100.}. So far, such attempts at unification have been
unsuccessful. Grand Unified Theories failed to be falsified by
observations (except for the simplest model: the $SU(5)$ model of
Georgi and Glashow~\cite{Glashow}). No one has yet succeeded in
showing that at least one of the many possible vacuum states in
string theory succinctly and correctly describes the Standard
Model of particle physics with the gauge group $SU(3)\times
SU(2)\times U(1)$ and 3 generations, although there are several
published low-energy supergravity models which describe in $D=4$
dimensions extensions of the standard model, e.g. $SO(32)$,
$SO(10)$ and $E_8\times E_8$. Due to the large number of possible
vacuum states, there exist many gauge groups that could describe
low-energy string theory. There is also no knowledge of the energy
scale at which the necessary supersymmetry breaking of string
theory should occur. Recently, it has been argued that the
supersymmetry breaking scale should be large, perhaps of order
1000 TeV or even at the Plank mass scale $\sim 10^{19}$
GeV~\cite{Dimopolous}. The reasons for this have to do with
several difficulties associated with a low-energy supersymmetry
breaking scale $\sim 1$ TeV, e.g. the failure to observe a light
Higgsino and the fast proton decay associated with dimension-5
operators in MSSM. Even if supersymmetric particles are observed
at the large hadron collider (LHC), it would be argued that the
solution to the electroweak hierarchy problem and the cosmological
constant problem are unavoidable fine-tuning problems associated
with low-energy physics, and should be understood in the context
of the anthropic principle. Without an observation of
superpartners, the whole edifice of supersymmetry and superstrings
is to be seriously questioned.

To suggest that we should deviate fundamentally from the way
physics has been done over the past three-four centuries by
constructing a ``statistical theory of
theories''~\cite{Douglas,Banks}, due to the failure of string
theory to provide a unique, falsifiable framework, is providing
present day string theory with an aura of validity that has not
yet been established. There is presently no known concrete
prediction made by string theory that can be falsified, although
more than 25 years have been devoted to discovering such a
prediction \footnote{The prediction of the correct number of
degrees of freedom according to the Bekenstein-Hawking entropy law
for black holes only holds uniquely for extremal black holes,
which are physically un-realizable objects~\cite{Strominger}. This result
does not constitute a laboratory experiment or astronomical observation
and, therefore, does not comply with the standard requirements of
well-tested physics theories.}. Because of this lack of testability, it
would seem inappropriate at present to use the elegance of string theory
to justify its correctness \footnote{In the 19th century, the existence
and nature of atoms and molecules were a mystery. In order to understand
atoms, Lord Kelvin (William Thompson)~\cite{Kelvin} invented an elegant
mathematical solution to the problem based on knot theory and topology. He
speculated that electromagnetic forces were propagated as linear and
rotational strains in an elastic solid (``ether''), producing vortex atoms
which generated the field. He proposed that these atoms consisted of tiny
knotted strings, and the type of knot determined the type of atom. The
correct theory of chemical bonding that was supported by experiment was
much less elegant than Kelvin's imaginative mathematical theory. In recent
years, knot theory has become a subject of considerable interest to
mathematicians.}.

The need for a de Sitter or almost de Sitter universe to explain
the acceleration of the expansion of the universe has also led to
problems in string theory, and complicated potentials have been
invented to produce possible meta-stable string solutions that
support ``inflation''~\cite{Burgess} and a positive cosmological
constant~\cite{Linde}. In view of the special initial conditions
required by standard inflationary cosmology, namely, a peculiar flat
inflaton potential and sufficient initial homogeneity of spacetime,
inflation theory is described by an ``eternal'' inflation
scenario~\cite{Vilenkin,Linde2}. This scenario requires a multiverse or
anthropic principle basis for its development. In contrast, the
alternative solution of initial value problems of early universe
cosmology, provided by a bimetric gravity theory~\cite{Moffat} with a
variable speed of light, predicts a power spectrum in agreement with the
data that does not require fine-tuned potentials or a large initial vacuum
energy and homogeneous universe. It postulates a large speed of light or a
small speed of gravitons in the early universe, depending upon which
metric frame is used to solve the cosmology, and there is no need for a
multiverse or the anthropic principle.

String theory does not appear to be
able to solve the ``cosmological constant problem'' using a ``monovacuum''.
We cannot, of course, exclude the possibility that some future form of
string theory will solve this problem, but in the present form of the
theory the possibility of doing so appears to face difficulties, even when
mechanisms to break supersymmetry are invoked, e.g. employing flux tubes
and D-branes. These mechanisms must use largely unknown non-perturbative
physics to break the string supersymmetry. This supersymmetry breaking is
essential to remove the massless moduli fields that disagree with
experiment.

The very large number of vacuum states that could possibly support
a small positive cosmological constant has led to the suggestion
that there can exist a large or even infinite number of universes
all with different values of $\Lambda$, some of which could
contain a vacuum density consistent with observations. After all,
if you take a large enough sample of universes, then one can argue
that some of them surely must have the correct vacuum state that
supports a very small $\Lambda$ and would ``make life possible."
As has been argued recently~\cite{Smolin2}, the fact that we
observe galaxies in {\it our} universe and that the cosmological
constant $\Lambda$ has to be smaller than some given value to
allow for the formation of galaxies~\cite{Weinberg2}, is simply an
observational fact and any statistical arguments that claim to
``explain'' this fact are irrelevant. It has also be pointed out
that we can conceive of universes in which both the cosmological
constant and the matter density are much higher and yet galaxies
can form~\cite{Aguirre}. Adherents of the anthropic principle
disagree with this conclusion, arguing that the anthropic
principle scenario and its consequences for the cosmological
constant problem is serious science~\cite{Vilenkin2}. There are
presently no compelling reasons to believe that the anthropic
principle can lead to a falsification of the string
landscape/topography scenario.

There have been times in the past when it seemed difficult and,
indeed, maybe impossible to explain some experimental observation,
e.g. the puzzle as to why the Sun could keep shining when its fuel
should have been spent a long time ago. This led to speculations
that if it did not continue to shine, then life on our planet
could not exist and therefore some ``anthropic'' principle must be
postulated to account for this dilemma. Another example is that it
took almost 50 years from the discovery of superconductivity by
Kammerling-Onnes in 1911 to the understanding of this phenomenon
by Bardeen, Cooper and Schrieffer~\cite{Cooper}. The lack of a
solution of the problem was due to the continuing attempt to use
na{\"i}ve perturbation theory when, all along, a non-perturbative
attack on the problem was required. Perhaps after many years of
trying to solve the problem of superconductivity, some physicists
began to feel that there was no solution, and that life as we know it
could not exist, if the phenomenon of superconductivity was not explained
and that only an ``anthropic principle'' argument could save the day. As
far as we know, this need was not overtly publicized by any of the
theorists who failed to understand the superconductivity phenomenon
\footnote{Recently, a possible resolution of the cosmological constant
problem was proposed based on the idea that the vacuum is unstable in the
presence of a gravitational field, causing fermion condensates to produce
a non-perturbative vacuum energy gap that leads to an exponential
suppression of the cosmological constant in the early
universe~\cite{Alexander}.}.

Given all of these arguments for and against the need for an
anthropic principle or multiverse solution to physical theory, we
shall nonetheless in the following pursue a statistical
interpretation of the string theory multivacuum problem. We can
conceive of many or even an infinite number of moduli
super-potentials that describe the possible vacua of string theory
as random fields. In Section 2, we shall adapt modified versions
of well-known methods for analyzing stochastic processes to string
theory to construct a Langevin equation for the string states,
such that the supersymmetric string vacua undergo a deformation
due to the stochastic fluctuations of the moduli potentials. In
Section 3, we derive a Focker-Planck equation for the string
probability distributions describing super-potentials
\footnote{Stochastic probability methods have been applied to
Einstein's general relativity theory~\cite{Moffat2} and the problem
of the cosmological constant~\cite{Moffat3,Sorkin}.}, while in
Section 4, these methods are applied to supergravity models for
the moduli K\"ahler potentials. In Section 5, we use the
stochastic fluctuations to attempt a solution of the cosmological
constant problem in string theory, and in Section 6 we end with
conclusions.

By following a statistical approach to string theory and its
vacua, we abandon a complete deterministic approach to string
theory as far as the solutions of the theory and its vacuum states
are concerned. Perhaps, we can regard this purely statistical
approach to the problem as an interim stage of the unification of
forces and quantum gravity and hope that in the future a more
unique and deterministic formulation of string theory will be
discovered, allowing for calculations using a unique vacuum state.

\section{Stochastic String Theory}

In string/M-theory the space of solutions is controlled by the
moduli space of supersymmetric vacua. The varying dynamical moduli
fields determine the size and shape parameters of compact internal
spaces that are required by 4-dimensional string theory. In a
supersymmetric theory these moduli fields are massless scalar
fields and cannot describe the real world. The continuum of
supersymmetric moduli fields has an exact super-particle
degeneracy and the associated cosmological constant $\Lambda$ is
zero. Therefore, we must seek non-supersymmetric moduli field
solutions and the number of these solutions is so huge that we may
as well say that it is infinite. The local minima of the
potentials $V(\phi)$ correspond to the possible values of the
cosmological constant and describe the topographical values
associated with valleys, while the peaks surrounding the extremal
valleys are the mountain tops describing saddle points or maximum
values of the moduli field solutions for the potentials. The
maximum extremal values of the potentials are meta-stable points
in the topography and if the local minima are absolute minima,
then the ``true'' vacuum states are stable. The simplest solution
of 10/11 string/M- theory is a flat Minkowski space manifested by
the massless moduli fields. No such massless particles have been
observed and we must modify the shape of this flat topography and
fix the values of the moduli by minimizing the potential energy.
If the minima are positive, then this indicates that the solutions
are described by de Sitter or near de Sitter solutions. It is
claimed that such solutions have been found by an intricate and
complicated combination of string techniques, including
D-brane/anti-brane configurations and fluxes of fields that are
higher-dimensional generalizations of electromagnetic field
sources~\cite{Linde}.

Let us consider string stochastic processes. The potential energy
$V(\phi)$ for the moduli fields $\phi$ satisfies the equation
\begin{equation}
\partial_{\phi}V(\phi)-f(\phi)=0,
\end{equation}
where $\partial_{\phi}=\partial V(\phi)/\partial{\phi}$ and we
adopt a real representation for the moduli fields $\phi$. The
$\phi$ symbolically denotes the complete set of supermoduli fields
$\phi_i$ associated with a string model and $f$ denotes the vector set
$f_i$. For a flat Minkowski vacuum solution $\langle V(\phi)\rangle=\langle
f(\phi)\rangle=0$, otherwise, for supersymmetric $AdS$ and $dS$ solutions
$\langle V(\phi)\rangle$ and $\langle f(\phi)\rangle$ do not in general
vanish. The infinite number of solutions for $V(\phi)$ will be described
by a stochastic process with random variables $F(\phi)$. The
Langevin~\cite{Langevin} string potential equation then reads
\begin{equation}
\label{Langevinpot}
\partial_{\phi}f(\phi)=-\sigma f(\phi)+F(\phi),
\end{equation}
where $\sigma$ is a constant. We assume that the stochastic
fluctuations of the superpotentials $V(\phi)$ are described by
Gaussian fluctuations and the stochastic variables $F(\phi)$
satisfy
\begin{equation}
\label{Fcorrelations} \langle F(\phi)\rangle=0,\quad \langle
F(\phi)F(\psi)\rangle=C\delta(\phi-\psi),
\end{equation}
where $C$ is a constant.

Solving for $f$ gives
\begin{equation}
\label{fequation} f=f_{0}\exp(-\sigma\phi)+\int_0^{\phi}
d\psi\exp(-\sigma(\phi-\psi))F(\psi),
\end{equation}
where we have cast (\ref{Langevinpot}) into an integral form using
either the Ito or Stratonovich~\cite{Ito} integral calculus. By
using (\ref{Fcorrelations}) we obtain
\begin{equation}
\langle f^2(\phi)\rangle
=f_0^2\exp(-2\sigma\phi)+\frac{C}{2\sigma}(1-\exp(-2\sigma\phi)).
\end{equation}
For large $\phi$ we shall make this equal $3E_s/m$, where $E_s$ is
the string energy scale and $m$ is the mean moduli field mass. It
follows that
\begin{equation}
\langle F(\phi)F(\psi)\rangle =\frac{6E_s}{m}\sigma
\delta(\phi-\psi).
\end{equation}

Integrating Eq.(\ref{fequation}) we obtain
\begin{equation}
V=V_0+\frac{V_0}{\sigma}(1-\exp(-\sigma\phi))+\int^{\phi}_0d\psi
\int^{\psi}_0d\chi\exp(-\sigma(\psi-\chi))F(\chi),
\end{equation}
where $V_0=V(0)$. This yields the mean square displacement
\begin{equation}
\langle(V(\phi)-V_0)^2\rangle
=\frac{V_0^2}{\sigma^2}(1-\exp(-\sigma\phi))^2
$$ $$
+\frac{3E_s}{m\sigma^2}(2\sigma\phi-3+4\exp(-\sigma\phi)-\exp(-2\sigma\phi)).
\end{equation}
For large $\phi$ this gives
\begin{equation}
\langle(V(\phi)-V_0)^2\rangle=\frac{6E_s}{m\sigma}\phi.
\end{equation}

\section{String Diffusion Focker-Planck Equation}

Let us introduce the probability distribution function
$p(V(\phi),f(\phi))$, which determines the probability $P$ to have
$V(\phi)$ and $f(\phi)$ for a given $\phi$:
\begin{equation}
P(V(\phi),f(\phi))=p(V(\phi),f(\phi),\phi)dV(\phi)df(\phi).
\end{equation}
If the fluctuations are small during the shifts $\Delta V(\phi)$
and $\Delta f(\phi)$, then we can assume that $V(\phi)$ and
$f(\phi)$ do not change significantly. We can then assume that the
string state at $V(\phi)+\Delta V(\phi)$ and $f(\phi)+\Delta
f(\phi)$ only depends on the string state at the values  $V(\phi)$
and $f(\phi)$, whereby the string state is Markovian.

Let us denote by the variable $u$ the pair
$u(\phi)=(V(\phi),f(\phi))$. We introduce the string transfer
function $\Psi(V,f,\Delta V,\Delta f,\Delta\phi)$ which describes
the probability to go from string state 1 to string state 2. The
evolution of the string Markov state depends on the transfer
function and the initial string state. We have
\begin{equation}
\int d\Delta\phi\Psi(u,\Delta u,\Delta\phi)=1,
\end{equation}
yielding a unit probability.

The stochastic string Kolmogorov equation is~\cite{Oksendal}:
\begin{equation}
p(u,\phi+\Delta\phi)=\int d\Delta up(u-\Delta u,\phi)\Psi(u-\Delta
u,\Delta u,\Delta\phi).
\end{equation}
For small values of $\Delta\phi$ we can expand $p(u)$:
\begin{equation}
p(u,\phi+\Delta \phi)=p(u,\phi)+\partial_{\phi}p\Delta\phi+...
\end{equation} and obtain the Focker-Planck equation (FPE)~\cite{Oksendal}:
\begin{equation}
\partial_{\phi}p(u,\phi)
=-\partial_u\biggl[f(u)p(u,\phi)\biggr]+\frac{1}{2}\partial^2_u[F^2(u)p(u,\phi)].
\end{equation}
In general, the probability evolution of the string system in
terms of the $V(\phi), f(\phi)$ ``phase'' space is described by
the moments $\langle(F(\phi))^n\rangle$. However, for our string
Focker-Planck equation, we solve the evolution only in terms of
the first and second moments.

A useful solution of the FPEs can be obtained for stationary
random $u(\phi)$ processes. We assume that a string system
subjected to $V(\phi)$ and $f(\phi)$ fluctuations can be described by a
stationary behavior of the probability distribution density $p$. This means
that the string system can form a state for which the probability density
$p_S(u)$ has a shape that does not change as $u$ develops. The
sample paths $u$ will in general not satisfy a steady state value $u_S$,
so that the $V$ and $f$ variables continue to fluctuate. However, these
fluctuations are such that $u$ and $u+\Delta u$ have the same probability
density, $p_S(u)$.

The stationary solution $p_S$ of the FPE satisfies
\begin{equation}
\partial_{\phi}p(u,\phi)+\partial_uI(u,\phi)=0,
\end{equation}
where
\begin{equation}
I(u,\phi)=f(u)p(u,\phi)-\partial_u[F^2(u)p(u,\phi)].
\end{equation}
The stationary FPE is then given by
\begin{equation}
\label{statsolution}
\partial_uI_{S}(u)=0,
\end{equation}
which implies that $I_{S}(u)={\rm constant}$.

The solution of the stationary FPE (\ref{statsolution}) reads
\begin{equation}
p_S(u)=\frac{C}{F^2(u)}\exp\biggl(2\int^u
dq\frac{f(q)}{F^2(q)}\biggr)
$$ $$
-\frac{1}{F^2(u)}I\int^udy\exp\biggl(2\int^u_y
dq\frac{f(q)}{F^2(q)}\biggr),
\end{equation}
 where $C$ denotes a normalization constant. When there is no flow
 of probability out of the string state, then for suitable boundary conditions,
 $I=0$. We then obtain
 \begin{equation}
 p_S(u)=\frac{C}{F^2(u)}\exp\biggl(2\int^u
 dq\frac{f(q)}{F^2(q)}\biggr).
 \end{equation}
 Here, we have defined the stochastic integrals by using the Ito
 prescription~\cite{Ito} which leads to a consistent calculus for
 random Gaussian fluctuations. In this case, the normalization
 constant $C$ is given by
 \begin{equation}
 C^{-1}=\int_{u_1}^{u_2}du\frac{1}{F^2(u)}
 \exp\biggl(2\int^udq\frac{f(q)}{F^2(q)}\biggr)
< \infty.
\end{equation}

\section{Low-energy Supergravity and Stochastic K\"ahler Potential Fluctuations}

Kachru et al.~\cite{Linde} (KKLT) have shown how geometric moduli
can be stabilized in IIB orientifold or F-theory compactifications
with ${\cal N}=1$ supersymmetry in $D=4$ dimensions. They begin
with type IIB string compactification with NS and RR 3-form
fluxes, H and F, respectively, through the three-cycles of a
Calabi-Yau manifold $M$. This yields a classical
superpotential~\cite{Gukov}:
\begin{equation}
{\tilde W}=\int_M\Omega\wedge G,
\end{equation}
where $G=F-\tau H$ and $\tau$ is the axion-dilaton. In
supergravity theory the effective 4-dimensional scalar potential
is given by~\cite{Bagger}
\begin{equation}
V=\exp(K)(G^{i\bar j}D_i WD_{\bar j}{\overline W}-3\vert
W\vert^2),
\end{equation}
where we have chosen units for which $M^{*2}_P=1$ and
\begin{equation}
D_i=\partial_iW+(\partial_jK)W.
\end{equation}
Moreover,
\begin{equation}
G_{i{\bar j}}=\partial_i\overline\partial_{\bar j}K
\end{equation}
is the K\"ahler metric obtained from the classical K\"ahler
potential
\begin{equation}
K=-\log(\int_M\Omega\wedge
\bar\Omega)-\log(\tau-\bar\tau)-2\log(\int_MJ^3).
\end{equation}
Here, $J$ is the form and $\Omega$ is the holomorphic 3-form on
$M$. The indices $i,j$ run over all the scalar fields, including
the complex structure moduli, $z_j$, the dilaton $\tau$, and the
complex K\"ahler moduli, $\rho_k=b_k+i\sigma_k$. The $b_k$ denote an
axion charge arising from the RR 4-form and $\sigma_k$ is the
4-cycle volume. In terms of the $t^i$ which measure areas of
2-cycles the classical volume is
\begin{equation}
{\cal V}=\int_MJ^3=\frac{1}{6}\kappa_{ijk}t^it^jt^k,
\end{equation}
and
\begin{equation}
\sigma_k=\partial_{t_k}{\cal V}=\frac{1}{2}\kappa_{ijk}t^it^j.
\end{equation}

The origin of the supergravity action can be either type IIB
string theory or the heterotic string theory. Type IIB string
theory is compactified on a Calabi-Yau manifold with K\"ahler
moduli. In general the number of moduli is large, but in the
literature usually only one or two moduli are analyzed and the
dilaton and the classical complex structure moduli have been
integrated out~\cite{Douglas,Polchinski,Alwis,Berglund,Frey}.

The equation
\begin{equation}
D_iW\equiv \partial_iW+(\partial_iK)W=0
\end{equation}
determines the supersymmetric vacua. The cosmological constant
becomes
\begin{equation}
\label{superlambda} \Lambda_{\rm susy}=-3\exp(K)\vert W\vert^2,
\end{equation}
corresponding to AdS or Minkowski spacetimes, and the latter hold
for $D_iW=W=0$. Choices of fluxes $G$ determine the $z_j$ and the
$\tau$. Moreover, when the superpotential is independent of
$\rho_k$, then the $3\vert W\vert^2$ cancels yielding a no-scale
potential $V$. This means that the stabilized supersymmetric
potential is independent of the K\"ahler moduli. KKLT suggested
two mechanism for stabilizing the K\"ahler moduli~\cite{Linde}:
(1) brane instantons and (2) gaugino condensates. Such
contributions are included as non-perturbative $W_{\rm np}$ effects in the
superpotential. Recently, Brustein and de
Alwis~\cite{Alwis} and Balasubramanian and
Berglund~\cite{Berglund} showed that stringy corrections to the
K\"ahler potential can give rise to non-supersymmetric vacua,
including metastable de Sitter spacetimes. They also claim that
the cosmological constant problem can be solved by a 1-loop
correction to the cosmological constant such that the scale of
supersymmetry breaking takes a realistic value. However, they are
still faced with the {\it discretuum} of values of possible $W$
and suggest that the anthropic principle should be used as a
selective principle or, failing this, that the cosmological
constant problem is an inevitable problem of fine-tuning.

We shall treat the ``lifting'' of the K\"ahler moduli to de Sitter
spacetime by our stochastic method. We postulate the SDE:
\begin{equation}
\label{LangevinW}
\partial_iW=f_i(K,W)+h_i(K,W),
\end{equation}
where
\begin{equation}
f_i(K,W)=-(\partial_iK)W,
\end{equation}
and $h_i$ is a stochastic variable which satisfies
\begin{equation}
\langle h_i(K,W)\rangle=0.
\end{equation}
We can cast this into an integral form
\begin{equation}
W=W^{0}+g^{ij}\int dY_if_j(Y)+g^{ij}\int dY_ih_j(Y),
\end{equation}
where $g^{ij}$ is a moduli field metric.

Let us denote by $w$ the absolute value of the superpotential,
$w=\vert W\vert$, and by $\Phi$ the absolute value of the moduli
variables, $\Phi=\vert\phi\vert$. For homogeneous Markov
processes, the FPE becomes
\begin{equation}
\partial_ip(w,\Phi_i)=-\partial_w[{\cal D}_i(w)p(w,\Phi_i)]
+\frac{1}{2}\partial^2_w[(H^2(w))_ip(w,\Phi_i)],
\end{equation}
where now $\partial_i=\partial_{\Phi_i}$, and the real drift
${\cal D}_i(w)$ and the real diffusion $H_i(w)$ are independent of
$\Phi_i$ and $H^2=g^{kl}H_kH_l$.

The real stationary probability density solution, $p_S$, is given
by
\begin{equation}
\partial_{\Phi} p_S(w,\Phi)+\partial_wI(w,\Phi)=0,
\end{equation}
where $\partial_{\Phi}p=\partial p/\partial\Phi$ and we have for
convenience of notation suppressed the indices $i,j$ and
\begin{equation}
I(w)={\cal D}(w)p(w)-\partial_w[H^2(w)p(w)].
\end{equation}
The stationary solution to the equation
\begin{equation}
\partial_wI(w)=0
\end{equation}
is given by
\begin{equation}
p_S(w)=\frac{C}{H^2(w)}\exp\biggl(2\int^{w}dX {\frac{{\cal
D}(X)}{H^2(X)}}\biggr)
$$ $$
-\frac{1}{H^2(w)}I(w)\int^w dY\exp\biggl(2\int^Y dX {\frac{{\cal
D}(X)} {H^2(X)}}\biggr).
\end{equation}

For no flow of probability, $I(w)=0$, we obtain
\begin{equation}
\label{stationaryprob} p_S(w)= \frac{C}{ H^2(w)}\exp\biggl(2\int^w
dX{\frac{{\cal D}(X)}{H^2(X)}}\biggr),
\end{equation}
where
\begin{equation}
C^{-1}=\int^{w_2}_{w_1}dw \frac{1}{H^2(w)}\exp\biggl(2\int^wdX
{\frac{{\cal D}(X)}{H^2(X)}}\biggr).
\end{equation}

A generic feature of the solution for the stationary probability
density (\ref{stationaryprob}) is that for a vanishing stochastic
average $\langle \vert H(w)\vert^2\rangle\rightarrow 0$, the
probability density will have a maximum value at, say, $w_{\rm
\max}$. Thus, the minimum values of the potentials for
non-supersymmetric vacua will occur {\it in the neighborhood of
the superpotential value $w_{\rm max}$} for small values of $H(w)$.

\section{The Cosmological Constant and Stochastic Fluctuations}

We shall assume that the stochastic supersymmetric breaking fluctuations
contribute to the ``lifting'' of the supersymmetric value of the
cosmological constant given by (\ref{superlambda}), so that the effective
cosmological constant is given by
\begin{equation}
\Lambda^2=\Lambda_{\rm susy}^2+\Lambda^2_{\rm fluct}+\Lambda_{\rm np}^2,
\end{equation}
where
\begin{equation}
\Lambda^2_{\rm fluct}=\langle V^2(\phi)\rangle,
\end{equation}
and $\Lambda_{\rm np}$ denotes the contributions to the effective
cosmological constant due to stringy non-perturbative effects.

 Quantum
``stringy'' corrections to $\Lambda_{\rm susy}$ will break supersymmetry
and ``hide'' superpartners from observation. The scale of the
non-perturbative supersymmetry breaking will occur at energies $E_{\rm sb}
> $ 1 TeV. We assume that the stochastic fluctuations that break
supersymmetry dominate the probabilistic determination of $\Lambda$, and in
particular that this is true at much lower energies, $E_{\rm fluct}\sim
10^{-3}$ eV, for small values of the stochastic fluctuations.  Moreover,
the magnitude of the value of $\Lambda_{\rm susy}$ must be small,
corresponding to small values of $K$ and $W$ in (\ref{superlambda}). Then,
it is possible to obtain a small positive value for the effective
cosmological constant $\Lambda$, corresponding to a vacuum energy density
$\rho_{\rm vac}\sim (10^{-3}\,eV)^4$, at the peak value of $\langle
V^2(\phi)\rangle$.

The actual supersymmetric breaking energy scale $E_{\rm sb}$ could
be at a large energy, $E_{\rm sb}\gg$ 1 Tev, but the stringy
corrections would not significantly affect the ''lifting`` of
$\Lambda_{\rm susy}$, whereas the stochastic fluctuations could
become significant already at the low-energy scale $E_{\rm
fluct}\sim 10^{-3}$ eV where the vacuum energy density $\rho_{\rm
vac}$ is found to give a positive de Sitter value of $\Lambda$.
Since the value of $\Lambda$ fluctuates as the universe expands, it is
possible to explain the ``coincidence'' of the onset of the acceleration of
the expansion of the universe in the present epoch.

\section{Conclusions}

We have adopted the point of view that if string theory and its
supergravity low-energy limit do correctly describe nature, then
the string topography should be described by a stochastic
probabilistic theory. We could interpret the different possible
values of the cosmological constant as being associated with
different universes, and the most likely value determined by our
stochastic probability theory is the one describing our universe.
However, we do not have to postulate a multiverse picture and its
accompanying anthropic principle. The many possible
extremal values of the superpotential in string theory are
described by a random variable and a non-vanishing correlation
function, formed from the stochastic random variable describing
the superpotentials as functions of the moduli fields $\phi$. The
most likely value of the superpotential is determined by a
Focker-Planck equation for the superpotential probability
distribution. Thus, a probabilistic depiction of string theory is
interpreted as a statistical solution of the string topography
{\it within our universe} and, as in quantum mechanics, a parallel
universes interpretation is a possible philosophical
interpretation of the world that can be avoided as a
non-falsifiable paradigm.

Our approach is different from the one described by the anthropic
principle, in that we postulate an ensemble of superpotentials and
classically solve for the most likely potential by means of
stochastic probability theory, because we do not have enough
information at present to solve string theory and obtain a unique
one vacuum solution. Thus, we say that we live in one universe and
we adopt a statistical approach to solving the superpotentials and
finding the most likely value for the vacuum state in our
universe. For the anthropic principle, on the other hand, each of
the string superpotentials and vacuum states that make up the
ensemble of universes represents a real result of an observation
in one of the universes, and only by saying that one or another of
the possible choices can sustain life can we decide which is
associated with our universe.

It is possible that we shall never discover a way to describe
string theory vacua by a monovacuum, leading to a
conventional quantum field theory solution to the cosmological
constant problem. Perhaps, only a statistical probability
description of the string topography will lead to a string
description of quantum gravity and particle physics.
\vskip 0.2
true in {\bf Acknowledgments}
\vskip 0.1 true in

This work was supported by the Natural Sciences and Engineering
Research Council of Canada. I thank Martin Green for helpful discussions.

\end{document}